\documentstyle[12pt]{article}
\title{Polygamma theory, the Li/Keiper constants, and validity of the
Riemann Hypothesis} 
\author{Mark W. Coffey\\
Department of Physics\\
Colorado School of Mines\\
Golden, CO  80401\\
(Received $\mbox{~~~~~~~~~~~~~~~~~~~~~~~~~~~~~~~2005}$)}
\date{March 6, 2005}
\begin{document}
\maketitle
\baselineskip=25 pt
\begin{abstract}

The Riemann hypothesis is equivalent to the Li criterion governing a 
sequence of real constants $\{\lambda_k\}_{k=1}^\infty$, that are certain 
logarithmic derivatives of the Riemann xi function evaluated at unity.
We investigate a related set of constants $c_n$, $n=1,2, \ldots$, 
showing in detail that the leading behaviour $(1/2) \ln n$
of $\lambda_n/n$ is absent in $c_n$.  Additional results are presented,
including a novel explicit representation of $c_n$ in terms of the Stieltjes
constants $\gamma_j$.  We conjecture as to the large-$n$ behaviour of $c_n$.
Should this conjecture hold, validity of the Riemann hypothesis would follow.


\end{abstract}
 
\vspace{.25cm}
\baselineskip=15pt
\centerline{\bf Key words and phrases}
\medskip 

\noindent
Li/Keiper constants, Riemann zeta function, Riemann xi function, logarithmic 
derivatives, Riemann hypothesis, Li criterion, Laurent expansion, Stieltjes 
constants

 
\baselineskip=25pt
\pagebreak
\medskip
\centerline{\bf Introduction}
\medskip


The Riemann hypothesis is equivalent to the Li criterion governing the 
sequence of real constants $\{\lambda_k\}_{k=1}^\infty$, that are certain 
logarithmic derivatives of the Riemann xi function evaluated at unity.
This equivalence results 
from a necessary and sufficient condition that the logarithmic derivative 
of the function $\xi[1/(1-z)]$ be analytic in the unit disk, where $\xi$ is 
the Riemann xi function.  The Li equivalence \cite{li}
states that a necessary and sufficient condition for the nontrivial zeros of
the Riemann zeta function to lie on the critical line Re $s=1/2$ is that 
$\{\lambda_k\}_{k=1}^\infty$ is nonnegative for every integer $k$.

This paper is a further contribution to our research program to characterize
the Li (Keiper \cite{keiper}) constants \cite{li,li2}.  
We have previously rederived \cite{coffey,coffey03} an arithmetic formula 
\cite{bombieri} for these constants, and described how it could be used to
estimate them.  Elsewhere, among several other results, we have examined
summatory properties of the Li and Stieltjes constants, and investigated
the $\eta_j$ coefficients appearing in the logarithmic derivative of the
zeta function about $s=1$ \cite{coffey04}.  In particular, a key feature of 
the sequence $\{\eta_j\}_{j=0}^\infty$ is now known:  it possesses strict sign alternation \cite{coffey04}.  

In this paper, we investigate a related set of constants \cite{wdsmith}
$c_n$, $n=1,2, \ldots$, that might be thought of as reduced Li/Keiper 
constants.  We show in detail that the leading behaviour $(1/2) \ln n$
of $\lambda_n/n$ is absent in $c_n$.  What remains in $c_n$ is a direct
manifestation of fundamental properties of the zeta function.  Thus,
$c_n$ can be variously interpreted as reflecting the nontrivial zeros, or
the $\eta_j$ constants.  
We present additional analytic results, including an explicit representation
of $c_n$ in terms of the Stieltjes constants $\gamma_k$.  We conjecture on
the precise order of $c_n$ in $n$, we comment on the nature of the logarithmic
derivative of the zeta function, and we briefly discuss possible
interpretations of some of our results.


The Gamma function is important in the theory of the Riemann zeta function--
for instance it is needed to complete $\zeta$ to the $\xi$ function.  The
Gamma function figures prominently in the functional equation for the zeta
function, thereby largely determining the location of the trivial zeros and 
other analytic properties.  Hence the digamma function appears in the
logarithmic derivative of the xi function, and higher derivatives introduce
the polygamma functions $\psi^{(j)}$ \cite{andrews}.  
Portions of the theory of this family 
of functions are very important in much of this paper.  We have also made 
extensive use of the properties of $\psi^{(j)}$ in previous works 
\cite{coffey,coffey03,coffey04}.

Our approach is explicit and very much in the spirit of constructivistic
mathematics.  Indeed, our work may be much more explicit than what would have
been thought possible just a few years ago.

From improved numerical calculation to height $T \simeq 2.38 \times 10^{12}$ 
\cite{gourdon}, it is now known that at least the first ten trillion complex 
zeros of the zeta function lie on the critical line.  For our purposes, this
effectively ensures that approximately the first $10^{26}$ $\lambda_k$'s are 
nonnegative, and this fact may have significant implications for our investigations.
For instance, it may very well turn out that working asymptotically in $k$
will suffice in our research program.  

The Li equivalence is by itself a qualitative reformulation of the Riemann
hypothesis.  The Riemann hypothesis does not of itself dictate the exact
nature of the Li/Keiper constants.  In fact, one can easily formulate
conjectures on the nature and order of the Li/Keiper constants that are then 
stronger than the Riemann hypothesis.  These observations indicate that the 
Riemann hypothesis may be verifiable without knowing the optimal order or other 
properties of the Li/Keiper constants that would more fully characterize them.

It is possible to use our approach also in pursuit 
of confirmation of the extended and generalized Riemann hypotheses.
The corresponding $\lambda$ constants have been defined for Dirichlet and
Hecke L-functions and other zeta functions \cite{li2}, and the same leading
behaviour $O(j\ln j)$ has been found \cite{coffey03}.  Our attention here
is strictly with the classical zeta function.  

\medskip
\centerline{\bf Preliminary Relations}

We first recall some notation, introduce some definitions, and present an
important Lemma for subsequent developments. We introduce the function
\cite{wdsmith}
$$F(z)=\ln \left[{z \over {1-z}}\zeta\left({1 \over {1-z}}\right)\right ],
\eqno(1)$$
whose analyticity in the unit disc $|z| < 1$ in the complex plane is 
equivalent to the Riemann hypothesis.  Then, should the power series
$$F(z)=\sum_{n=0}^\infty c_n z^n \eqno(2)$$
converge for $|z| < 1$, the Riemann hypothesis follows.  In short order,
one verifies that the $c_n$ are real constants, $c_0=0$, and $c_1=\gamma$,
the Euler constant.  Therefore, we may write
$$F(z)=\gamma z + \sum_{n=2}^\infty c_n z^n.  \eqno(3)$$

This paper investigates the behaviour of the constants $c_n$.  The importance
of this subject is clear:  from any subexponential bound on them the Riemann
hypothesis follows \cite{bombieri,wdsmith}.  
Indeed, in this paper we additionally conjecture the true order of the 
constants $|c_n|$, such that this conjecture is stronger than the Riemann 
hypothesis itself.    

Suppose we knew that $F(z)$ is analytic and univalent within the unit disc.
Then the function $F(z)/\gamma$, satisfying $F(0)/\gamma=0$ and $F'(0)/\gamma=1$,
is schlicht, fulfills the conditions for the Bieberbach conjecture \cite{debranges}
to hold, and thus we would have $|c_n| \leq \gamma n$ for all $n=1, 2, \ldots$.  
This shows the self consistency of the complex analysis involved.  In fact, any
direct application to the Bieberbach conjecture is thwarted due to the essential
singularity in $F$, making it highly non-univalent.    

Related to a prefactor in Eq. (1), $k(z)=z/(1-z)^2$ is the Koebe function, and
$k_\alpha(z)=z/(1-\alpha z)^2$ with $|\alpha|=1$ are rotations of it.  These are 
the only functions for which equality holds in the conclusion of the 
Bierberbach conjecture.  The history of the Bieberbach conjecture shows that it
is easier to obtain results about the logarithmic coefficients of a univalent
function rather than for the coefficients of the function itself \cite{koepf},
and the approach of Smith \cite{wdsmith} seems to fit within this framework.
  
The classical Laurent expansion of the Riemann zeta function about the unique pole
at $s=1$ introduces the Stieltjes constants $\gamma_k$ 
\cite{ivic,scref,matsuoka,matsuoka2}, with $\gamma_0=\gamma$.  We have
$$\zeta(s)={1 \over {s-1}}+ \sum_{n=0}^\infty {{(-1)^n} \over {n!}}\gamma_n
(s-1)^n, \eqno(4)$$
where the Stieltjes constants can be written in the form
$$\gamma_k=\lim_{N \to \infty}\left (\sum_{m=1}^N {1 \over m}\ln^k m-
{{\ln^{k+1} N} \over {k+1}}\right ). \eqno(5)$$
and several other forms have been given \cite{scref}. It is clear that the 
$c_n$'s are multinomials in the Stieltjes constants, and that $c_n$ contains 
terms $-(-1)^n\gamma^n/n$ and $-(-1)^n\gamma_{n-1}/(n-1)!$.  
Indeed, we are able to write much more, giving, for instance,
an explicit formula for $c_n$ in terms of the $\gamma_k$'s.  Once again, we
may therefore observe that sufficient estimation of the Stieltjes constants
would provide verification of the Riemann hypothesis.      

We discuss these connections with the Stieltjes constants in a later section.
For the moment, we simply point out that Appendix A contains explicit formulae
for the first few $c_n$'s in terms of them.  

The conformal map introducing the Li/Keiper constants and the coefficients 
$c_n$ is a natural one.  This map $z=1-1/s \leftrightarrow s=1/(1-z)$ takes the 
right-half plane Re $s > 1/2$ to the interior of the unit circle in the complex $z$-plane.  Just this sort of mapping arises in the theory of finite fields,
whose zeta functions have zeros on a circle in the complex plane 
\cite{weil2,ireland,bombieri74,vandam}. 
We recall that Weil proved the Riemann hypothesis holds for 
nonsingular curves over a finite field \cite{weil2}, while Deligne established
the validity of Weil's conjectures for generalized hypersurfaces that may
include intersections of hypersurfaces \cite{ireland}.  

We first present the explicit connection between the Li/Keiper constants and the 
constants $c_n$.  We have
\newline{{\bf Lemma 1}}
$${\lambda_n \over n} = c_n + {1 \over n}-{1 \over 2}\ln \pi + d_n, ~~~~~~~~
n \geq 1, \eqno(6)$$
where $d_n$ is the coefficient of $z^n$ of the function $\ln \Gamma[1/2(1-z)]$,
and $\Gamma$ is the Gamma function.  That is,
$$d_n = {1 \over {n!}}{d^n \over {dz^n}} \ln \Gamma \left[{1 \over {2(1-z)}}
\right ]_{z=0}.  \eqno(7)$$
We will present not only $d_n$, but the derivatives themselves in this equation.
We then estimate $d_n$ in $n$ and demonstrate that the leading behaviour of
$\lambda_n/n$ in Eq. (6) exactly cancels it.  This is highly supportive of a
decrease of $|c_n|$ with $n$, and of the conclusion that the Riemann hypothesis
should hold.  Since we have previously conjectured as to the subdominant
behaviour of the Li constants \cite{coffey03,coffey04}, we thereby have an
immediate conjecture for $|c_n|$.  

Before proving the Lemma and going on to expressions for $d_n$, we give some 
brief background on the Li (or Keiper) constants.  The function $\xi$ is 
determined from $\zeta$ by the relation 
\cite{davenport,edwards,ivic,karatsuba,titch,riemann}
$$\xi(s)={s \over 2}(s-1)\pi^{-s/2}\Gamma\left({s \over 2}\right) \zeta(s), 
\eqno(8)$$
and satisfies the functional equation $\xi(s)=\xi(1-s)$.  
The sequence $\{\lambda_n\}_{n=1}^\infty$ is defined by
$$\lambda_n = {1 \over {(n-1)!}}{d^n \over {ds^n}}[s^{n-1}\ln \xi(s)]_{s=1}.
\eqno(9)$$
The $\lambda_j$'s are 
connected to sums over the nontrivial zeros of $\zeta(s)$ by way of 
\cite{keiper,li}
$$\lambda_n=\sum_\rho\left [1-\left(1-{1 \over \rho}\right)^n \right ],
\eqno(10)$$ 
and
$$\lambda_1=-{{\ln \pi} \over 2}+{\gamma \over 2}+1 -\ln 2. \eqno(11)$$                  

In the representation \cite{bombieri,coffey,coffey03,coffey04}
$${\lambda_n \over n}={1 \over n}S_1(n)+{1 \over n}S_2(n)-{1 \over 2}
(\gamma +\ln \pi + 2\ln 2), \eqno(12)$$
the sum
$$S_1(n) \equiv \sum_{m=2}^n (-1)^m {n \choose m} (1-2^{-m})\zeta(m), ~~~~~~
n \geq 2,  \eqno(13)$$
has been characterized \cite{coffey03}:
$${n \over 2}\ln n + (\gamma - 1){n \over 2} +{1 \over 2} \leq S_1(n) 
\leq {n \over 2}\ln n+(\gamma+1){n \over 2} - {1 \over 2}. \eqno(14)$$ 
Further bounds on $S_1(n)$ have been developed by applying Euler-Maclaurin
summation to all orders \cite{coffey03}.  

For the sum
$$S_2(n) \equiv -\sum_{m=1}^n {n \choose m} \eta_{m-1}, \eqno(15)$$
the constants $\eta_j$ can be written as
$$\eta_k={(-1)^k \over {k!}} \lim_{N \to \infty}\left (\sum_{m=1}^N {1 \over 
m} \Lambda(m)\ln^k m - {{\ln^{k+1} N} \over {k+1}}\right ), \eqno(16)$$
and $\Lambda$ is the von Mangoldt function 
\cite{edwards,ivic,karatsuba,titch,riemann}, 
such that $\Lambda(k)=\ln p$ when $k$ is a power of a prime $p$ and 
$\Lambda(k)=0$ otherwise.
The constants $\eta_j$ enter the expansion around $s=1$ of the logarithmic 
derivative of the zeta function, 
$${{\zeta'(s)} \over {\zeta(s)}} = -{1 \over {s-1}}-\sum_{p=0}^\infty \eta_p 
(s-1)^p, ~~~~~~|s-1| < 3, \eqno(17)$$ 
and the corresponding Dirichlet series valid for Re $s>1$ is
$${{\zeta'(s)} \over {\zeta(s)}}=-\sum_{n=1}^\infty {{\Lambda(n)} \over n^s}.
\eqno(18)$$
The constants $\eta_j$, with $\eta_0=-\gamma$, have been written explicitly in
terms of the Stieltjes constants \cite{coffey04,maslanka}, a point on which we
return later. Additionally, we recently proved the strict sign alternation of the
sequence $\{\eta_j\}_{j=0}^\infty$ \cite{coffey04}.

Now that the sum $S_2(n)$ has been introduced, we may present the exact relation
{\newline {\bf Lemma 2}}
$${{S_2(n)} \over n} = c_n.  \eqno(19)$$
For, from Eq.\ (2) it follows that $c_n =(1/n!)(d^n/dz^n) F(z)|_{z=0}$, and from
Eqs.\ (8) and (9) we have
$$S_2(n)={1 \over {(n-1)!}}{d^n \over {ds^n}}\left[s^{n-1} \ln[(s-1)\zeta(s)
\right]]_{s=1}$$
$$=\sum_{m=1}^n {n \choose m}{1 \over {(m-1)!}} {d^m \over {ds^m}}\ln \left
[(s-1)\zeta(s)\right]_{s=1}.  \eqno(20)$$
Under the mapping $s(z)=1/(1-z)$, derivatives transform as
$d/dz=s^2 d/ds$, and the Lemma follows.

\pagebreak
\centerline{\bf Proof of Lemma 1 and expressions for $d_n$}
\medskip

From Eqs. (1) and (8) we have
$$\ln \xi\left({1 \over {1-z}}\right)=F(z)-\ln(1-z)+{1 \over {2(z-1)}}\ln \pi
+\ln \Gamma\left[{1 \over {2(1-z)}}\right ].  \eqno(21)$$
Since we have \cite{li,li2}
$$\ln \xi\left({1 \over {1-z}}\right)=\sum_{n=1}^\infty {\lambda_n \over n}z^n,
\eqno(22)$$
the expansion of Eq. (21) in powers of $z$ readily yields Eq. (6).  (For the
latter equation, we have used the convention $\xi_{Li}(z)=2\xi(z)$, such 
that $\xi_{Li}(0)=1$, in place of $\xi(z)$.  Otherwise, Eq. (6) will have 
another minor term $-\ln 2$.)

From Lemmas 1 and 2 it follows that
{\newline {\bf Corollary}}
$${{S_1} \over n}-{\gamma \over 2}={1 \over n}+d_n.  \eqno(23)$$


We next have
{\newline {\bf Lemma 3}}
$${d^n \over {dz^n}} \ln \Gamma \left[{1 \over {2(1-z)}}\right ]
= {1 \over 2}{d^{n-1} \over {dz^{n-1}}}{1 \over {(1-z)^2}}\psi\left[{1 \over
{2(1-z)}}\right]$$
$$={1 \over 2}\sum_{j=0}^{n-1} {{n-1} \choose j} {{(n-j)!} \over {(1-z)^{n-j+1}}}
{d^j \over {dz^j}} \psi\left[{1 \over {2(1-z)}}\right]$$
$$={1 \over 2}\left\{{{n!} \over {(1-z)^{n+1}}} \psi\left[{1 \over {2(1-z)}}\right]
+\sum_{j=1}^{n-1} {{n-1} \choose j} {{(n-j)!} \over {(1-z)^{n+1}}}
\sum_{\ell=1}^j {j \choose \ell} {{(j-1)!} \over {(\ell-1)!}}{1 \over {2^\ell
(1-z)^\ell}}\psi^{(\ell)}\left[{1 \over {2(1-z)}}\right] \right \}, 
\eqno(24)$$
where $\psi$ is the digamma function and $\psi^{(j)}$ is the polygamma function.

We mention two proofs of this Lemma.  The key ingredient is knowing how to
write the successive derivatives of the digamma factor.  This can be 
accomplished by applying the Faa di Bruno formula for the $n$th derivative of 
a composite function.  We have
$${d^n \over {dz^n}} \psi\left[{1 \over {2(1-z)}}\right]
={1 \over {(1-z)^n}}\sum_{j=1}^n {n \choose j} {{(n-1)!} \over {(j-1)!}}{1 \over {2^j
(1-z)^j}}\psi^{(j)}\left[{1 \over {2(1-z)}}\right]. 
\eqno(25)$$
This equation is just a slight extension of a formula for the derivatives of a
function $\theta(1/x)$ \cite{cofpla}.  
Equation (25), when used with the product rule, completes the Lemma.

Another method can be based upon the expansion \cite{nbs,andrews}
$$\Gamma(z)={1 \over z}\exp\left[-\gamma z+\sum_{k=2}^\infty {{(-1)^k \zeta(k)} \over
k} z^k\right], \eqno(26)$$\
giving
$$\ln \Gamma\left[{1 \over {2(1-z)}}\right]=\ln 2-\sum_{n=1}^\infty {z^n \over n}
-{\gamma \over 2}\sum_{n=0}^\infty z^n+\sum_{k=2}^\infty {{(-1)^k\zeta(k)} \over 
{k2^k}} \left (\sum_{j=0}^\infty z^j\right )^k.  \eqno(27)$$
Expanding the powers of power series on the right side returns us to the Lemma.
Afterall, we have the relation 
$$\psi^{(j)}\left({1 \over 2}\right)=(-1)^{j+1}j!(2^{j+1}-1)\zeta(j+1),  
~~~~~~j\geq 1,  \eqno(28)$$
and this is very helpful in rewriting the constants $d_n$ below.

From Eqs. (25) and (28) we have an exact reformulation of digamma derivatives of
interest:
{\newline {\bf Lemma 4}}
$${d^n \over {dz^n}} \psi\left[{1 \over {2(1-z)}}\right]_{z=0}
=n!\sum_{m=1}^\infty {1 \over m^2}\left[2\left(1-{1\over m}
\right)^{n-1}-{1 \over 2}\left(1-{1 \over {2m}}\right)^{n-1}\right ].  \eqno(29)$$
This Lemma follows by setting $z=0$ in Eq. (25), inserting Eq. (28), using the
Dirichlet series for the zeta function, and reordering the double sum.  The use
of a derivative relation of a binomial sum completes the work.

We are in position to write compact, yet exact, expressions for the constants
$d_n$:
{\newline {\bf Lemma 5}}
$$d_n={1 \over 2}\psi\left({1 \over 2}\right)+{1 \over {2n}}\sum_{j=1}^{n-1}
(n-j) \sum_{m=1}^\infty {1 \over m^2}\left[2\left(1-{1 \over m}\right)^{j-1}
-{1 \over 2}\left(1-{1 \over {2m}}\right )^{j-1}\right]$$
$$={1 \over 2}\psi\left({1 \over 2}\right)+{1 \over {2n}}\sum_{m=1}^\infty
\left[2\left(1-{1 \over m}\right)^n-2\left(1-{1 \over {2m}}\right )^n
+{n \over m} \right], ~~~~~~n \geq 1, \eqno(30)$$
where $\psi(1/2)/2=-\gamma/2-\ln 2$.
The first line of the Lemma follows from the combination of the results of
Lemmas 2 and 3, and the second line follows from application of finite 
geometric series.

\medskip
\centerline{\bf High order approximation for the constants $d_n$}
\medskip

There are many ways in which to obtain highly accurate approximations to 
$d_n$ for large values of $n$.  The upshot is 
{\newline {\bf Lemma 6}}
$$\ln \Gamma\left[{1 \over {2(1-z)}}\right ] \simeq {1 \over 2}\ln \pi
+{1 \over 2}\sum_{j=1}^\infty \left[\psi(j)+\gamma-\ln 2 -1
\right ] z^j .  \eqno(31)$$
That is, for $j >> 1$ we have  
$$d_j = {1 \over 2}\left[\ln j -{1 \over {2j}}-{1 \over {12j^2}}+\gamma-\ln 2-1+O\left({1 \over j^4} \right)\right ].  \eqno(32)$$
In connection with Eq. (31), we recall the value of the digamma function at
integer argument in terms of harmonic numbers $H_n$:  $\psi(n)=H_{n-1}-\gamma$.

We indicate a couple of approaches for obtaining Lemma 6.  One is based upon
using the integral corresponding to the summation on the second line of Eq. (30).
It turns out that this integral, $I_1(n)$, was extensively studied in Appendix
A of Ref. \cite{coffey03}, and we have taken over the results.

In another method, we apply Euler-Maclaurin summation to the sum over $m$
on the first line of Eq. (30).  In doing so, we put
$$f(m) \equiv {1 \over m^2}\left[2\left(1-{1\over m}\right)^{n-1}
-{1 \over 2}\left(1-{1 \over {2m}}\right)^{n-1}\right ],  \eqno(33)$$
such that $f(1)=-2^{-n}$, $f(\infty)=0$, and we have the elementary integral
$$\int_1^\infty f(k) dk = {{1+2^{-n}} \over n}.  \eqno(34)$$
Therefore we obtain
$${d^n \over {dz^n}} \psi\left[{1 \over {2(1-z)}}\right]_{z=0}
\simeq n!\left ({{1+2^{-n}} \over n} - 2^{-n-1}\right).  \eqno(35)$$
Then the sum over $j$ can be performed in Eq. (30).  In either approach,
we discard any terms in the final result that are exponentially small in $n$,
such as $2^{1-n}/n$.

Of note, all terms in $d_n$ beyond the leading logarithmic dependence are in
terms of integral powers of $1/n$--there is no algebraic dependence upon $n$.

\medskip
\centerline{\bf Relations and formulae for $c_n$}
\medskip

With the aid of Cauchy's integral formula, the $c_n$'s can be written as
$$c_n={1 \over {2\pi i}}\int_C {{F(z)} \over z^{n+1}} dz, \eqno(36)$$
where $C$ is a simple closed contour about the origin, and this can
serve as the basis of a numerical method \cite{wdsmith}.  If $C$ is a circle
of radius $r$ about the origin, then $c_n = r^{-n}\int_0^1 F(re^{2\pi i \phi})
e^{-2\pi i n \phi}d\phi$, and this invites the use of fast Fourier transform
for evaluation.
 
The combination of the result of Lemma 5 with Eqs. (6), (12), and (14) 
shows that $c_n = S_2(n)/n + O(1/n)$, and Lemma 2 gives the
strengthening to $c_n=S_2(n)/n$.
Since we have previously conjectured that $|S_2(n)|=O(n^{1/2+\varepsilon})$
for $\varepsilon >0$, we have
{\newline {\bf Conjecture}}
$$|c_n| = O\left({1 \over n^{1/2-\varepsilon}}\right ), \eqno(37)$$
where $\varepsilon >0$ but is otherwise arbitrary.  That is, we anticipate
that the magnitudes $|c_n|$ decrease nearly as the square root of $n$ for
large $n$.  In Figure 1 we compare such a decrease with available numerical
evidence \cite{wdsmith,maslanka,coffey03}.  Figure 1 contains a 
semilogarithmic plot of $|c_n|$ versus $n$, together with a curve 
corresponding to $6/\pi^2 \sqrt{n}$.
For this limited set, after a few initial values, the latter curve appears to 
provide a consistent upper bound.  In light of the von Koch result on the
Riemann hypothesis that $\psi(x) = x + O(x^{1/2}\ln^2 x)$ \cite{ingham}, where 
$\psi$ is the Chebyshev function, we suspect that the optimal order of $|c_n|$ 
is very close to $O(\ln n/n^{1/2})$.  
Figure 2 shows an example plot of $c_n$ versus $n$, 
that illustrates the oscillatory behaviour of these constants.  In addition,
Smith has now numerically confirmed our conjecture for the first approximately
$10^5$ values of $c_n$ \cite{wdsmith}.
%

In Figure 3 we have plotted the differences $\delta_n=c_n^2-c_{n-1}c_{n+1}$
versus $n$, that appear to support a decrease in $|c_n|$ with increasing
$n$.  In addition, the behaviour may indicate a correlation in the sign or other
properties of the $c_n$'s.  Figure 4 plots the magnitude of the discrete
Fourier transform applied to this sequence.  This plot indicates underlying
structure.

Our conjecture suggests that it may be worthwhile to study in detail the
properties of the particular polylogarithm 
$$L_{1/2}(z) \equiv \sum_{n=1}^\infty {z^n \over n^{1/2}}, \eqno(38)$$
such that $L_{1/2}(-1)= (\sqrt{2}-1)\zeta(1/2)$.

Previously, we obtained an expression for the Li/Keiper constants explicitly in 
terms of the Stieltjes constants \cite{coffey04}.  We recall this and 
related results \cite{coffey04}:
{\newline{\bf Theorem}}
$$\lambda_n = 1-{n \over 2}(\ln \pi+2\ln 2-\gamma)+S_1(n)-\sum_{j=2}^n(-1)^j
{n \choose j} j\sum_{h=1}^j {1 \over h} \sum_{\stackrel{j_1 \geq 0, ...,
j_h \geq 0}{j_1+\cdots+j_h=j-h}}  
\prod_{b=1}^h {{\gamma_{j_b}} \over {j_b!}}, ~~~~~~n \geq 2,  \eqno(39)$$
$$S_2(n)=n\gamma-\sum_{j=2}^n(-1)^j{n \choose j} j\sum_{h=1}^j {1 \over h} 
\sum_{\stackrel{j_1 \geq 0, ...,j_h \geq 0}{j_1+\cdots+j_h=j-h}} 
\prod_{b=1}^h {{\gamma_{j_b}} \over {j_b!}}, ~~~~~~n \geq 2,  \eqno(40)$$
and  
{\newline{\bf Theorem}}
$$\eta_{k-1} = (-1)^k k\sum_{h=1}^k {1 \over h}
\sum_{\stackrel{j_1 \geq 0, ...,j_h \geq 0}{j_1+\cdots+j_h=k-h}} 
\prod_{b=1}^h {{\gamma_{j_b}} \over {j_b!}}, ~~~~~~k \geq 2.  \eqno(41)$$

From Lemma 2 we obtain the exact relation
{\newline {\bf Theorem}} 
$$c_n=\gamma - {1 \over n}
\sum_{j=2}^n(-1)^j{n \choose j} j\sum_{h=1}^j {1 \over h} 
\sum_{\stackrel{j_1 \geq 0, ...,j_h \geq 0}{j_1+\cdots+j_h=j-h}} 
\prod_{b=1}^h {{\gamma_{j_b}} \over {j_b!}}, ~~~~~~n \geq 2.  \eqno(42)$$
On the right side of Eq. (42), the constrained sum over the indices
$j_\ell$ means that we have a partition of $k-h$ over the nonnegative 
integers.  All such partitions are considered, meaning that their order
does not matter.  The number of such partitions is ${n-1 \choose h-1}$
in $\eta_{n-1}$ or $c_n$.   

From Eq. (4) we have
$${z \over {1-z}}\zeta\left({1 \over {1-z}}\right)=1+\sum_{n=0}^\infty
{{(-1)^n} \over {n!}} \gamma_n \left({z \over {1-z}}\right)^{n+1}.  \eqno(43)$$
Then from the definition (1) and performing various expansions, we have
$$F(z)=-\sum_{n=1}^\infty {{(-1)^n} \over n}\left[\sum_{j=0}^\infty {{(-1)^j} 
\over {j!}}\gamma_j z^{j+1} \left(\sum_{m=0}^\infty z^m\right)^{j+1}\right]^n.
\eqno(44)$$
Carrying out the expansion in powers of $z$ in this equation must necessarily
return us to Eq. (42) for the coefficients $c_n$.

From the Hadamard product formula for the zeta function (e.g., \cite{titch}),
$$\zeta(s)={{\exp(\ln 2\pi -1-\gamma/2)s} \over {2(s-1)\Gamma(s/2+1)}}
\prod_\rho \left(1-{s \over \rho}\right )e^{s/\rho}, \eqno(45)$$
we obtain
$$F(z)={{\ln 2\pi-1-\gamma/2} \over {1-z}} -\ln 2-\ln \Gamma\left[{{3-2z} \over 
{2(1-z)}}\right ]+\sum_\rho\left\{\ln \left[1-{1 \over {\rho(1-z)}}\right ]
+{1 \over {\rho(1-z)}}\right \}.  \eqno(46)$$
Since by Eq. (10), $\sum_\rho 1/\rho \equiv \sigma_1 = \lambda_1$, we have
$$F(z)={{\ln \pi} \over {2(1-z)}} -\ln 2-\ln \Gamma\left[{{3-2z} \over 
{2(1-z)}}\right ]+\sum_\rho\ln \left[1-{1 \over {\rho(1-z)}}\right ].
\eqno(47)$$
The value $F(0)=\sum_\rho \ln[(\rho-1)/\rho]=0$ obtains because the sum
over all the complex zeros of $\zeta$ contains the pairs of $\rho$ with
$1-\rho$.

From the functional equation in the form $\zeta(z)=\pi^{z-1}2^z\Gamma(1-z)
\zeta(1-z)\sin(\pi z/2)$, we obtain
$$F\left({1 \over z}\right)=F(z)-\ln z+{{\ln (-\pi)} \over {z-1}}+{z \over {z-1}}
\ln 2+\ln \Gamma\left({1 \over {1-z}}\right)+\ln \sin\left[{\pi \over 2}{z \over
{(z-1})}\right ].  \eqno(48)$$
This equation should be very useful in obtaining results on the boundedness of
the $c_n$'s.

\medskip
\centerline{\bf Discussion of the logarithmic derivative of $\zeta$}
\medskip

This function has proved to be central in analytic number theory.  Here
we recall some known results and relate them to Eqs. (17), (18), and others.


We have (e.g., \cite{riemann,titch}) in terms of the prime counting function 
$\pi(x)$
$$\ln \zeta(s) = s\int_2^\infty {{\pi(x)} \over {x(x^s-1)}} dx, ~~~~~~\mbox{Re}
~ s >1.  \eqno(49)$$
Then
$${{\zeta'(s)} \over {\zeta(s)}} =\int_2^\infty {{\pi(x)} \over {x(x^s-1)}}dx
-s\int_2^\infty {{x^{s-1} \pi(x) \ln x} \over{(x^s-1)^2}}dx, ~~~~~~\mbox{Re}~s >1. 
\eqno(50)$$

Since the function $\pi$ has steps, these induce changes in the coefficients
$\eta_j$, hence in $S_2(n)$ or the $c_n$ constants.  In the common region
of validity Re $s >1$ $\cap$ 
$|s-1|<3$, we have from Eqs. (17) and (49)
$$[(s-1)+1]\int_2^\infty {{\pi(x)} \over {x(x^s-1)}} dx=-\ln(s-1)-\sum_{p=1}^\infty
{\eta_{p-1} \over p} (s-1)^p.  \eqno(51)$$
That is, we could write for instance
$$-\ln(s-1)-\sum_{p=1}^\infty {\eta_{p-1} \over p} (s-1)^p
=[(s-1)+1]\int_2^\infty dx{{\pi(x)} \over x}\left\{{1 \over {x-1}}-{{x\ln x} \over
{(x-1)^2}}(s-1)\right.$$
$$\left. +\left[-{x \over 2}{{\ln^2 x} \over {(x-1)^2}}+{{x^2 \ln^2 x} \over
{(x-1)^3}}\right](s-1)^2 + O[(s-1)^3] \right \},  \eqno(52)$$
where of course by the prime number theorem $\pi(x) \sim x/\ln x$ as $x \to \infty$.
This equation in powers of $s-1$ gives in principle an integral representation
for each of the coefficients $\eta_j$.  

We mention an important occurrence of the logarithmic derivative in numerical
analysis.  The reciprocal of this function is key in the classical Newton
iteration for root finding, bringing in connections with discrete dynamical systems.  
Then one seeks the attracting fixed points of the associated Newtonian mapping.
Therefore, from this point of view it is not unexpected that the logarithmic
derivative should play an important role in determining zeros.

\medskip
\centerline{\bf Summary and Brief Discussion}
\medskip





By way of Lemma 2, or equivalently from the use of the theory of polygamma
functions, we have shown that the constants $c_n$ of Eqs. (2) and (3) omit
the leading growth $(1/2) \ln n$ of $\lambda_n/n$.  We have presented 
additional analytic arguments, a conjecture, and partial numerical results
that point to the decrease of $|c_n|$ with $n$.  We have pointed out the limited
possibility of directly applying the Bieberbach conjecture because the function $F$
of Eq. (1) is not univalent within the unit disc.

The quantity $S_2(n)$ is formed as the binomial sum of the alternating $\eta_j$
values of Eq. (17).  The latter is a correlated sequence.  For, we have 
previously exhibited \cite{coffey04} the explicit summatory relation imposed upon 
the $\eta$'s by the functional equation of either the zeta or xi functions.
This relation implies that a given $\eta_j$ is connected to all the other
values $\eta_{j+1}$, $\eta_{j+2}$, $\dots$.


In Appendix B we call out an integral representation of the alternating zeta
function that may permit a joining of probabilisitic interpretation of the
zeta function with Krein spectral shift functions.  In turn, this may provide
a useful link between Hardy space theory and inverse scattering theory.
Though fairly independent of the approach of this paper, we believe it may be worth
pointing this out to other investigators.


The importance of an explicit formula for $S_2(n)$ or $c_n$ should not
be overlooked.  For instance, in principle, only improved estimation of the
Stieltjes constants prevents verification of the Riemann hypothesis by
way of either the Li criterion \cite{li} or by way of Criterion (c) of Ref.
\cite{bombieri}.  Concerning the magnitudes $|c_n|$, any subexponential bound
would serve to verify the Riemann hypothesis.

\pagebreak
\centerline{\bf Acknowledgements}
\medskip
I thank W. van Dam for useful discussion.  I thank K. Ma\'{s}lanka for
useful correspondence and the numerical values of $S_2(n)$ used for the figures.
I thank W. D. Smith for useful correspondence.
This work was partially supported by a SPARC grant from Regis University.

\pagebreak
\medskip
\centerline{\bf Figure Captions}

FIG. 1.  In this semilogarithmic plot, the upper curve corresponds to values
of $6/\pi^2 \sqrt{n}$ versus $n$, and the lower to values of $|c_n|$ versus $n$.

FIG. 2.  Plot of $c_n$ versus $n$.

FIG. 3.  Plot of the differences $\delta_n=c_n^2-c_{n-1}c_{n+1}$ versus $n$.

FIG. 4.  Plot of the magnitude of the discrete Fourier transform of the $c_n$
sequence.

\pagebreak
\centerline{\bf Appendix A:  Examples of $c_n$ in terms of the Stieltjes constants}
\medskip

Here, $\gamma$ is the Euler constant and $\gamma_k$ are the Stieltjes constants
appearing in Eq. (4).  We have
$$c_1=\gamma, ~~~~~~~~c_2=\gamma-\gamma^2/2-\gamma_1, \eqno(A.1)$$
$$c_3=\gamma-\gamma^2+{1 \over 3}\gamma^3-2\gamma_1+\gamma\gamma_1+{1 \over 2}
\gamma_2, \eqno(A.2)$$
$$c_4=\gamma^3-{1 \over 4}\gamma^4-{1 \over 2}\gamma^2(3+2\gamma_1)
+\gamma(1+3\gamma_1-{1 \over 2}\gamma_2)+{1 \over 6}[-3\gamma_1(6+\gamma_1)
+9\gamma_2-\gamma_3], \eqno(A.3)$$
and
$$c_5=-\gamma^4+{1 \over 5}\gamma^5+\gamma^3(2+\gamma_1)+{1 \over 2}\gamma^2
(-4-8\gamma_1+\gamma_2)+\gamma[1+\gamma_1(6+\gamma_1)-2\gamma_2+{1 \over 6}
\gamma_3]$$
$$+{1 \over {24}}[72\gamma_2+12\gamma_1(-8-4\gamma_1+\gamma_2)
-16\gamma_3+\gamma_4].  \eqno(A.4)$$

The first few $d_k$'s are given by
$$d_0={1 \over 2}\ln \pi, \eqno(A.5)$$
$$d_1=-\gamma/2-\ln 2, \eqno(A.6)$$
$$d_2=-\gamma+{1 \over 8}\pi^2-2\ln 2, \eqno(A.7)$$
$$d_3=-3\gamma+{3 \over 4}\pi^2-6\ln 2-{7 \over 4} \zeta(3), \eqno(A.8)$$
$$d_4=-12\gamma+{9 \over 2}\pi^2 +{\pi^4 \over {16}}-24 \ln 2-21 \zeta(3),
\eqno(A.9)$$
and
$$d_5=-60\gamma+30\pi^2+{5 \over 4}\pi^4-120 \ln 2-210 \zeta(3)-{{93} \over 4}
\zeta(5).  \eqno(A.10)$$

\pagebreak
\centerline{\bf Appendix B:  The alternating zeta function in inverse spectral
theory}
\medskip

We first recall the alternating zeta function
$$\sum_{n=1}^\infty {{(-1)^{n-1}} \over n^s} = (1-2^{1-s})\zeta(s), ~~~~~~~
\mbox{Re}~ s > 0, ~~~~ s \neq 1, \eqno(B.1)$$
this being one of the many analytic continuations of the Dirichlet series
for the Riemann zeta function.  Without going into the details, it turns out
that this function can be written as an integral representation with the
Krein spectral shift function \cite{krishna} associated with the harmonic
oscillator Hamiltonian on the line, with a Dirichlet boundary condition at
the origin.  As a Corollary, one may write \cite{krishna}
$$(1-2^{1-s})\zeta(s) = s\int_0^\infty e^{-sx} \phi(x)dx, ~~~~~~~~~
\mbox{Re}~ s > 0, \eqno(B.2)$$
where
$$\phi(x)=\sum_{n=1}^\infty \chi_{[\ln(2n-1),\ln 2n]}(x), \eqno(B.3)$$
and $\chi$ is the characteristic function of an interval.  
%
More generally, if the Dirichlet boundary condition is enforced at any
other point $x$, a family of functions $\zeta(x,s)$ is generated.  
The points of discontinuity of $\zeta(x,s)$ satisfy a differential 
equation in $x$ called the Dubrovin equation.  This differential equation
gives a curve in the space of analytic functions with the alternating
zeta function divided by $s$ as the initial value \cite{krishna}.  

Now it is possible to represent the function \cite{ehm}
$$\eta(s) \equiv {{(s-1)} \over s^2} \zeta(s)=\int_0^\infty e^{-xs} \phi_1(x)dx, 
~~~~~~~ \mbox{Re}~ s > 0, \eqno(B.4)$$
with the real-valued function
$$\phi_1(x)=\sum_{1 \leq n \leq e^x} (1+ \ln n -x).  \eqno(B.5)$$
Then it is possible to introduce a family of probability densities with
$x \in [0,\infty)$ as $p_\sigma(x)=\phi_1(x)\exp(-\sigma x)/\eta(\sigma)$
for $\sigma >0$ \cite{ehm}.  The cumulants of $p_\sigma$ can be written 
either in terms of the Stieltjes constants or the Li/Keiper constants at
$\sigma=1$ \cite{coffey04,ehm}.  Comparing Eqs. (B.2) with (B.4) and (B.3)
with (B.5), it appears that it should be possible to combine a probabilistic
setting for the zeta function with an inverse spectral theory.  This point
of view offers a connection between quantum dynamics and stochastic
processes.

\pagebreak


\begin{thebibliography}{99}
\bibitem{nbs}M. Abramowitz and I. A. Stegun,
{Handbook of Mathematical Functions, Washington, National Bureau of Standards
(1964).}
\bibitem{andrews}G. E. Andrews, R. Askey, and R. Roy, 
{Special Functions, Cambridge University Press (1999).}
\bibitem{bombieri}E. Bombieri and J. C. Lagarias,
{Complements to Li's criterion for the Riemann hypothesis,
J. Number Theory {\bf 77}, 274-287 (1999).}
\bibitem{bombieri74}E. Bombieri,
{Counting points on curves over finite fields (d'apr\`{e}s S. A. Stepanov),
Sem. Bourbaki, Vol. 1972-73, Expos\'{e} 430.  Lecture Notes in Mathematics,
Vol. 383, 234-241, Springer (1974).}
\bibitem{coffey}M. W. Coffey,
{Relations and positivity results for derivatives of the Riemann $\xi$
function, J. Comput. Appl. Math., {\bf 166}, 525-534 (2004).}
\bibitem{coffey03}M. W. Coffey,
{Towards verification of the Riemann hypothesis, preprint, 2003.}
\bibitem{coffey04}M. W. Coffey,
{New results concerning power series expansions of the Riemann xi function and
the Li/Keiper constants, preprint, 2005.}
\bibitem{cofpla}M. W. Coffey,
{A set of identities for a theta function at unit argument, Phys. Lett. A,
{\bf 300}, 367-369 (2002).}
\bibitem{davenport}H. Davenport,
{Multiplicative Number Theory, Springer Verlag (2000).}
\bibitem{debranges}L. de Branges,
{A proof of the Bieberbach conjecture, Acta Math. {\bf 154}, 137-152 (1985).}
\bibitem{edwards}H. M. Edwards,
{Riemann's Zeta Function, Academic Press, New York (1974).}
\bibitem{ehm}W. Ehm,
{A family of probability densities related to the Riemann zeta function,
Contemp. Math. {\bf 287}, 63-74 (2001).}
\bibitem{gourdon}X. Gourdon,
{The $10^{13}$ first zeros of the Riemann zeta function and computations at 
very large height, preprint (2004).}
\bibitem{ingham}A. E. Ingham,
{The distribution of prime numbers, Cambridge University Press (1990).}
\bibitem{ireland}K. Ireland and M. Rosen,
{A classical introduction to modern number theory, Springer (1990).}
\bibitem{scref}M. I. Israilov,
{On the Laurent Decomposition of Riemann's zeta function,
Dokl. Akad. Nauk SSSR (Russian) {\bf 12}, 9 (1979); M. I. Israilov,
Trudy Mat. Inst. Steklova {\bf 158}, 98-104 (1981); G. H. Hardy, 
Note on Dr. Vacca's series for $\gamma$, Quart. J. Pure
Appl. Math. {\bf 43}, 215-216 (1912); J. C. Kluyver, On certain series of
Mr. Hardy, Quart. J. Pure Appl. Math. {\bf 50}, 185-192 (1927); W. E. 
Briggs, Some constants associated with the Riemann zeta-function, 
Mich. Math. J. {\bf 3}, 117-121 (1955); D. Mitrovi\'{c}, The signs of some
constants associated with the Riemann zeta function, Mich. Math. J. {\bf 9}, 
395-397 (1962).}
\bibitem{ivic}A. Ivi\'{c}, 
{The Riemann Zeta-Function, Wiley (1985).}
\bibitem{karatsuba}A. A. Karatsuba and S. M. Voronin,
{The Riemann Zeta-Function, Walter de Gruyter, New York (1992).}
\bibitem{keiper}J. B. Keiper,
{Power series expansions of Riemann's $\xi$ function, Math. Comp. {\bf 58}, 
765-773 (1992).}
\bibitem{koepf}W. Koepf,
{Power series, Bieberbach conjecture, and the de Branges and Weinstein functions,
ISSAC'03 (2003).}   
\bibitem{krishna}M. Krishna,
{$\xi$ - $\zeta$ relation, Proc. Indian Acad. Sci. {\bf 109}, 379-383 (1999).}
\bibitem{li}X.-J. Li,
{The positivity of a sequence of numbers and the Riemann hypothesis,
J. Number Th. {\bf 65}, 325-333 (1997).}
\bibitem{li2}X.-J. Li,
{Explicit formulas for Dirichlet and Hecke L-functions, Ill. J. Math.,  
{48}, 491-503 (2004).}
\bibitem{maslanka}K. Ma\'{s}lanka,
{Effective method of computing Li's coefficients and their properties, preprint
(2004); An explicit formula relating Stieltjes constants and Li's numbers,
preprint (2004), submitted to Exptl. Math.}
\bibitem{matsuoka}Y. Matsuoka,
{A note on the relation between generalized Euler constants and the zeros of the
Riemann zeta function, J. Fac. Educ. Shinshu Univ. {\bf 53}, 81-82 (1985);
A sequence associated with the zeros of the Riemann zeta function, Tsukuba J.
Math. {\bf 10}, 249-254 (1986).}
\bibitem{matsuoka2}Y. Matsuoka,
{Generalized Euler constants associated with the Riemann zeta function,
in:  Number Theory and Combinatorics, ed. by J. Akiyama et al., World 
Scientific (1985).  Concerning just Section 6 here, $t$ should read $T$ on
p. 293 (line 4), $\Gamma$ should read $T$ on the same line, and $+$ should
read $=$ on the next line.  On the next page, $\gamma_0^3$ should be
replaced with $\gamma_0^4$ in Example 1.}
\bibitem{riemann}B. Riemann,
{\"{U}ber die Anzahl der Primzahlen unter einer gegebenen Gr\"{o}sse, 
Monats. Preuss. Akad. Wiss., 671 (1859-1860).}
\bibitem{wdsmith}W. D. Smith,
{Cruel and unusual behavior of the Riemann zeta function, 
http://www.math.temple.edu/$\sim$wds/homepage/works.html, preprint, 1998; revised
2005}
\bibitem{titch}E. C. Titchmarsh,
{The Theory of the Riemann Zeta-Function, 2nd ed., Oxford University
Press, Oxford (1986).}
\bibitem{vandam}W. van Dam,
{Quantum computing and zeroes of zeta functions, quant-ph/0405081 (2004).}
\bibitem{weil2}A. Weil,
{Number of solutions of equations in a finite field, Bull. Am. Math. Soc. {\bf 55},
487-495 (1949); Courbes alg\'{e}briques et vari\'{e}t\'{e}s ab\'{e}liennes, 
Hermann (1971); Oeuvres Scientifiques--Collected Papers, 3 Vols., Springer
(1980).}  
\end{thebibliography}
\end{document}